\documentclass[twocolumn,showpacs,preprintnumbers,amsmath,amssymb,pre]{revtex4}

\usepackage{graphicx}
\usepackage{dcolumn}
\usepackage{bm}
\usepackage{enumerate}
\usepackage{color}

\begin{document}

\title{Diffusion of antibiotics through a biofilm in the presence of diffusion and absorption barriers}

\author{Tadeusz Koszto{\l}owicz}
\email{tadeusz.kosztolowicz@ujk.edu.pl}
\affiliation{Institute of Physics, Jan Kochanowski University, Uniwersytecka 7, 25-406 Kielce, Poland}
 
\author{Ralf Metzler}
\email{rmetzler@uni-potsdam.de}
\affiliation{Institute for Physics and Astronomy, University of Potsdam, D-14476, Potsdam-Golm, Germany}      

\date{\today}

\begin{abstract}

We propose a model of antibiotic diffusion through a bacterial biofilm when diffusion and/or absorption barriers develop in the biofilm. The idea of this model is: We deduce details of the diffusion process in a medium in which direct experimental study is difficult, based on probing diffusion in external regions. Since a biofilm has a gel-like consistency, we suppose that subdiffusion of particles in the biofilm may occur. To describe this process we use a fractional subdiffusion-absorption equation with an adjustable anomalous diffusion exponent. The boundary conditions at the boundaries of the biofilm are derived by means of a particle random walk model on a discrete lattice leading to an expression involving a fractional time derivative. We show that the temporal evolution of the total amount of substance that has diffused through the biofilm explicitly depends on whether there is antibiotic absorption in the biofilm. This fact is used to experimentally check for antibiotic absorption in the biofilm and if the biofilm parameters change over time. We propose a four-stage model of antibiotic diffusion in biofilm based on the mentioned above physical characteristics. The biological interpretation of the stages, in particular their relation with the bacterial defence mechanisms, is discussed. Theoretical results are compared with empirical results of ciprofloxacin diffusion through {\it Pseudomonas aeruginosa} biofilm, and ciprofloxacin and gentamicin diffusion through {\it Proteus mirabilis} biofilm. 

\end{abstract}

\maketitle

\section{Introduction\label{secI}}

We present a model of antibiotic diffusion through a bacterial biofilm in which absorption of antibiotic molecules can occur. Since a biofilm has a gel-like consistency, we suppose that subdiffusion of particles in the biofilm may occur. To describe this process we use a fractional subdiffusion-absorption equation. The model is of a general nature and can be used to study diffusion processes in media in which experimental diffusion investigations are difficult. The application of this model is based on the idea: {\it We can specify details of the diffusion process in a medium in which experimental diffusion study is difficult, based on diffusion properties observed in external regions}. Experimental methods based on the model are non--invasive to the biofilm. We show that the temporal evolution of the total amount of substance that has diffused through the medium explicitly depends on whether there is absorption of diffusing particles in the medium. We divide the process into different stages according to the following criteria: (a) whether there is absorption of diffusing particles in the medium or not, (b) whether the diffusion and absorption parameters are constant or change over time. The potential application of this model goes beyond the specific problem we use as a guiding example. Namely it is a generic model to deduce diffusion properties from the particle currents exchanged with the direct environment.

The model is used to experimentally check for antibiotic absorption in the biofilm and whether the biofilm parameters change over time. We define four stages in antibiotic diffusion-absorption process in a biofilm according to the criteria how points (a) and (b) are met. The division into stages is made according to physical, not biological criteria. However, determining the order of stages and its duration may help in the biological interpretation of antibiotic interaction with bacteria processes. We present possible criteria which of the biofilm defence mechanisms can be considered as dominant at each stage. However, this topic is still open and requires further research, as more such mechanisms are being discovered.

Bacterial biofilms play a key role in persistent infections. Bacteria in a biofilm develop increased resistance of antimicrobial agents. There are many ways to defend the bacteria against antibiotic molecules. Transport limitation is an important factor in the antimicrobial resistance of biofilm bacteria \cite{aot,mot,chambless,jacobs,stewart_1996}. One of the symptoms of bacterial defence against antibiotics is to slow down the diffusion and retain antibiotic molecules in the biofilm. Observation of antibiotic diffusion through a bacterial biofilm allows one to understand the physical and biological processes occurring in the biofilm. 

Models of antibiotic diffusion in the biofilm take into account specific changes in the biofilm resulting from the defence of bacteria against the antibiotic. To describe this process the normal diffusion or normal diffusion--reaction equations have been usually used \cite{stewart_1996,stewart_1994,stewart_2016,vrany,aristotelous,taherzadeh,beyenal,nichols,anguige,birnir,gade,acunto,balsa,klapper,wang}. Because the biofilm has a gel-like consistency, the movement of antibiotic molecules is rather strongly hindered. Therefore, as in gel--like media \cite{tk2005,tk2005a,nera,lieleg,ctwm,jlom,gbm,wong}, subdiffusion may occur in the biofilm. In this case, the subdiffusion--reaction equation with fractional time derivative is a convenient approach. 

One of the key problems is to find the boundary conditions at the biofilm boundary.
Particle random walk models on a discrete lattice are effective at deriving boundary conditions at the border between media. Some models assume that there is a point at the boundary between media at which the molecule must be stopped temporarily \cite{vankampen,lomholt,zaid,goychuk,korabel}. In another model, it is assumed that the molecule can jump across the border between the media without having to stop at the border \cite{kijhmt,tk2019}. In general, both models lead to different boundary conditions. In our considerations, we assume that a molecule that tries to get out of the biofilm can do it without having to stop at the edge of the biofilm. Therefore, in the following the latter model will be used to derive the boundary conditions.

A biofilm changes as a result of bacterial interaction with antibiotics. Bacterial defence mechanisms against antibiotics result in specific processes, such as absorption or slowing down of diffusion of antibiotic molecules in the biofilm, these processes may occur with varying intensity. We distinguish four stages of antibiotic diffusion in a biofilm. These stages are defined by the following criteria: (a) if there is absorption of antibiotic molecules in the biofilm or if absorption is absent, (b) if biofilm parameters are constant or if at least one parameter changes over time. 

Various experimental techniques are used to study the processes occurring in the biofilm in the presence of antibiotics, such as imaging microprocesses in biofilm, disk diffusion methods, chromatography methods etc. \cite{zhang,balouiri}. Another technique for measuring the effect of antibiotics on bacteria based on measuring the temporal evolution of the amount of a specifics antibiotic that has diffused through the biofilm $W_B$ has been shown in \cite{ar,ar1}. We will show that the function $W_B$ differs qualitatively for the stages mentioned earlier, which gives the opportunity to experimentally check in which stage the process is. As examples, we show that the theoretical function describes well empirical results of ciprofloxacin diffusion through {\it Pseudomonas aeruginosa} PAO1 biofilm, and ciprofloxacin and gentamicin diffusion through {\it Proteus mirabilis} O18 biofilm \cite{ar,ar1}.

\section{Antibiotic diffusion in a biofilm\label{secII}}

Bacteria exist mainly as planktonic bacteria and in biofilms. Biofilms are complex microbial communities of cells embedded into a matrix of self-produced extracellular polymeric substance. The organization of bacteria in biofilm helps in defending bacteria against antibiotics. Bacteria in biofilms have even 1000 times greater resistance to antibiotics compared to bacterial plankton. In a biofilm, bacteria have many different ways of defending against an antibiotic. The most often considered biofilm defence mechanisms are \cite{aot,mot,chambless}:
(i) the biofilm matrix may act as a diffusion barrier, 
(ii) microenvironments are created in which slower bacterial growth occurs. In these regions, the effect of the antibiotic is weakened, because the antibiotic act strongly mainly on fast-growing bacteria. Examples of this are regions where oxygen and nutrient access are reduced, 
(iii) the presence of persisters in biofilm. The persisters are small subpopulation of bacteria which weaken the effect of antibiotic, 
(iv) the resistance genes which regulate the biofilm defence mechanism. Which way of defence is dominant depends on both a biofilm and the specific antibiotic. In addition to the above, there are many other factors, such as some nontoxic colloidal particles \cite{lu} and increased extracellular polymeric substance production in older biofilms \cite{singla}, that increase the defence ability of bacteria against the action of antimicrobial molecules. Bacteria may also exchange DNA pieces and pass on successful mutations increasing the immune properties of the biofilm. Quorum sensing is a cell-to-cell communication phenomenon which affects the cell population density and regulates their behaviour. This phenomenon also influences the increase of biofilm resistance to the antibiotic \cite{aot,mot,kpcm}. 

As we mentioned earlier, models of antibiotic diffusion in a biofilm have been based mainly on normal diffusion or normal diffusion-reaction equations. In \cite{stewart_1994} the interaction of an antibiotic with the biofilm was modelled taking into account the antibiotic depletion process and reduced bacterial growth rates in biofilm. Normal diffusion-reaction equation with different reaction terms were considered in \cite{stewart_2016}. In both papers simple boundary conditions at the biofilm boundaries are assumed, namely, vanishing of the diffusion flux of the antibiotic or keeping a constant antibiotic concentration at the biofilm boundaries.
The diffusion--adsorption equation has been used to describe antibiotic diffusion in a {\it Pseudomonas aeruginosa} bioflim \cite{nichols}. This equation is equivalent to the normal diffusion equation with diffusion coefficient controlled by an adsorption parameter. Normal diffusion equations taking into account the absorption and desorption processes were used to model transport of ciprofloxacin and levofloxacin in {\it Pseudomonas aeruginosa} biofilms \cite{vrany}.
In addition to the diffusion of antibiotics, other factors affecting the biofilm have been included in the models, such as oxygen diffusion into biofilm \cite{gade}, influence of persister cells to antibiotic diffusion \cite{roberts}, and the quorum sensing phenomenon \cite{anguige,kpcm}. 

Here we present an alternative approach based on a fractional diffusion mechanism. We explicitly derive the corresponding boundary value problem involving a fractional time derivative. Our results are shown to be consistent with experimental observations in two different biofilm--forming species.

\section{Model\label{secIII}}

In this section, we present the system, the general assumptions adopted in the model, and the boundary conditions at the border between biofilm and normal--diffusion medium. 

\subsection{System\label{secIIIA}}

\begin{figure}[htb]
\centerline{%
\includegraphics[scale=0.45]{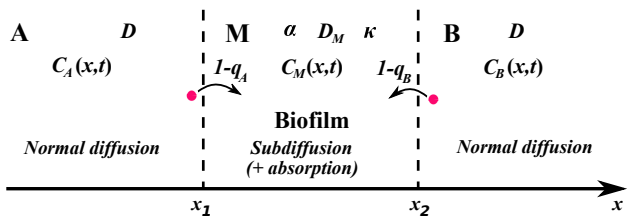}}
\caption{Schematic of the system. The biofilm separates two regions in which normal diffusion occurs, $D$ is the normal diffusion  coefficients in regions $A$ and $B$, $D_M$ is the subdiffusion coefficient, $\alpha$ is the subdiffusion parameter and $\kappa$ is the absorption coefficient in the biofilm, $q_A$ and $q_B$ are probabilities of stopping a diffusing particle by the biofilm boundaries.}
\label{fig1}
\end{figure}

Our considerations concern a three--dimensional system which is homogeneous in the plane perpendicular to the $x$ axis. Thus, later in this paper we treat this system as one--dimensional. We consider the system which is schematically presented in~Fig.~\ref{fig1}. 
The system consists of three parts: $A$, $(-\infty,x_1)$, and $B$, $(x_2,\infty)$, represent normal diffusion media, the middle part $M$, $(x_1,x_2)$, represents a biofilm. A molecule that attempts to jump from the media $A$ or $B$ to the biofilm can do it with probabilities $1-q_A$ and $1-q_B$, respectively. A molecule that tries to get out of the biofilm can do it without any hindrance.

\subsection{Assumptions\label{secIIIB}}

The model of diffusion of antibiotic molecules through a biofilm is based on the following assumptions:

(i) {\it There may be subdiffusion in the biofilm}. Subdiffusion is due to the complex structure of the medium, which makes diffusion of molecules very difficult \cite{mk,mk1}. Indeed, the polymeric structure connecting cells in a biofilm is similar to gels, e.g. aqueous agarose solution \cite{tk2005,tk2005a,kijhmt}. Moreover, similar to mucus, charge effects may came into play. In many cases diffusion in similar environments may be anomalous. We therefore base our description on subdiffusion of antibiotic molecules in a biofilm, although normal diffusion is included as a limiting case.

(ii) {\it Absorption of antibiotic molecules may occur in the biofilm}. Absorption is treated here as an irreversible reaction, the result of which is to switch off the antibiotic molecule from further action. The molecule can be invoked in a dense biofilm or it can interact with the bacterium.

(iii) {\it We use an approximation of a homogeneous biofilm}. We assume that the subdiffusion and absorption parameters in the biofilm do not depend on the spatial variable. This assumption has been often used in the models presented in the articles cited in the previous sections.

(iv) {\it The antibiotic molecule that attempts to jump from a diffusion medium to a biofilm can do it with a certain probability, and the molecule that tries to leave a biofilm will do it without any hindrances}. The problem of getting an antibiotic molecule inside the biofilm can be caused by biofilm defence mechanisms. Moreover, a molecule that tries to jump into a biofilm from an external diffusive medium has to hit one of the channels in the biofilm. A molecule that tries to get out of the biofilm does not encounter such obstacles. Although we use the approximation of a homogeneous biofilm, we assume that the probabilities of retaining diffusing molecules at biofilm surfaces $q_A$ and $q_B$ may be different. The motivation for this assumption is that the external concentrations of the antibiotic, which may be different at both biofilm boundaries, affect bacterial defence mechanisms at the boundaries. We also assume that the boundaries of the biofilm do not significantly change their position over time. 

(v) {\it Parameters of subdiffusion and/or absorption in the biofilm can change over time; in the considerations we use a `quasistatic approximation'.} It is supposed that the subdiffusion--absorption process in the biofilm is slow. Then, the solutions to the equation with parameters changing over time will be obtained in the following way. First, we will solve the equation with fixed parameters and then we will change the parameters into time-dependent functions.
This assumption is consistent with the concept of the stationary phase in the modelling of antibiotic diffusion in the biofilm \cite{aot,anderl}. 

\subsection{Equations\label{secIIIC}}

We assume that in parts $A$, $M$, and $B$ of the system the process is described by the following equations 
\begin{equation}\label{eq1}
  \frac{\partial C_A(x,t)}{\partial t}=D\frac{\partial^{2}C_A(x,t)}{\partial x^{2}},
\end{equation}
\begin{eqnarray}\label{eq2}
  \frac{\partial C_M(x,t)}{\partial t}=D_M\frac{\partial^{1-\alpha}}{\partial t^{1-\alpha}}\Bigg[\frac{\partial^{2}C_M(x,t)}{\partial x^{2}}
	-\kappa^2 C_M(x,t)\Bigg],
\end{eqnarray}
\begin{eqnarray}\label{eq3}
  \frac{\partial C_B(x,t)}{\partial t}=D\frac{\partial^{2}C_B(x,t)}{\partial x^{2}},
\end{eqnarray}
where $D_M$ has physical dimension $m^2/sec^\alpha$.
The Riemann--Liouville fractional derivative, which is present in Eq. (\ref{eq2}), is defined for $0<\beta<1$ as 
\begin{equation}\label{eq4}
  \frac{d^{\beta}f(t)}{dt^{\beta}}=\frac{1}{\Gamma(1-\beta)}\frac{d}{dt}\int_{0}^{t}dt'\frac{f(t')}{(t-t')^{\beta}}\;.
\end{equation}
The diffusive fluxes are defined as $J_{A,B}(x,t)=-D\partial C_{A,B}(x,t)/\partial x$ and $J_M(x,t)=-D_M(\partial^{1-\alpha}/\partial t^{1-\alpha})\partial C_M(x,t)/\partial x$.

For $\alpha=1$ we have normal diffusion whereas for $0<\alpha<1$ there is subdiffusion. The appearance of the fractional time derivative in the subdiffusion equation means that the process is non-Markovian with a long memory. In this case, according to the Continuous Time Random Walk model, the time distribution for the next jump of the molecule $\psi$ has a heavy tail, $\psi(t)\sim 1/t^{1+\alpha}$ when $t\rightarrow\infty$, which gives rise to an infinite characteristic sojourn time $\left\langle t\right\rangle$ \cite{mk}.

\subsection{Boundary conditions\label{secIIID}}

\begin{figure}[htb]
\centerline{%
\includegraphics[scale=0.45]{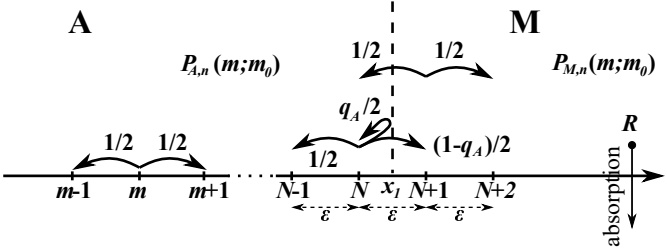}}
\caption{Random walk of a particle in a discrete system with one--sided fully permeable wall represented by the vertical line, more detailed description in the text.}
\label{fig2}
\end{figure}

It is essential to determine the boundary conditions at the boundaries of the biofilm.
In order to derive them we use the particle random walk model in a system with a one--sided fully permeable wall \cite{tk2019}. Within the model we assume that both variables, the particle position $m$ and time $n$, are discrete. Finally, we move to continuous variables $x$ and $t$. 
As an example, we derive the boundary conditions at $x_1$. Since the boundary conditions for normal diffusion and subdiffusion are local, for the sake of simplicity we assume that there is one partially permeable wall in the system located between sites $N$ and $N+1$, which corresponds to the biofilm boundary at $x_1$, see Fig. \ref{fig2}. 
The difference equations describing a random walk in this system are 
\begin{eqnarray}\label{eq5}
P_{A,n+1}(m;m_0)=\frac{1}{2}P_{A,n}(m-1;m_0)\\
+\frac{1}{2}P_{A,n}(m+1;m_0),\;\; m\leq N-1,\nonumber
\end{eqnarray}
\begin{eqnarray}\label{eq6}
P_{A,n+1}(N;m_0)=\frac{1}{2}P_{A,n}(N-1;m_0)\\
+\frac{1}{2}P_{M,n}(N+1;m_0)
+\frac{q_A}{2}P_{A,n}(N;m_0),\nonumber
\end{eqnarray}
\begin{eqnarray}\label{eq7}
P_{M,n+1}(N+1;m_0)=\frac{1-q_A}{2}P_{A,n}(N;m_0)\\
+\frac{1}{2}P_{M,n}(N+2;m_0)
-R P_{M,n}(N+1;m_0),\nonumber
\end{eqnarray}
\begin{eqnarray}\label{eq8}
P_{M,n+1}(m;m_0)=\frac{1}{2}P_{M,n}(m-1;m_0)\\
+\frac{1}{2}P_{M,n}(m+1;m_0)
-R P_{M,n}(m;m_0),\;\;m\geq N+2,\nonumber
\end{eqnarray}
where $P_{i,n}(m;m_0)$ is the probability to find the particle at site $m$ in region $i$ after $n$ steps, $m_0$ is the initial position of the particle, and $R$ is the probability of particle absorption in the medium $M$. The Green's functions for continuous time, in terms of the Laplace transform $\hat{P}(m,s)\equiv \mathcal{L}[P(m,t)]\equiv \int_0^\infty {\rm exp}(-st)P(m,t)dt$, is 
\begin{equation}\label{eq9}
\hat{P}_i(m,s;m_0)=\frac{1-\hat{\psi}_i(s)}{s}S_i(m,\hat{\psi}_i(s);m_0),
\end{equation}
where $S_i(m,z;m_0)=\sum_{n=0}^\infty z^n P_{i,n}(m;m_0)$ is the generating function,
and $\psi_i$ is the probability density of time which is needed for the particle to take its next step in the medium $i$. Moving from discrete to continuous spatial variable we use the following relations $x=\epsilon m$, $x_1=\epsilon N$, $x_0=\epsilon m_0$, and $\hat{P}(x,s;x_0)=\hat{P}(m,s;m_0)/\epsilon$, where $\epsilon$ is the distance between neighbouring sites. We then take the limit of small $\epsilon$. As it was shown in \cite{tk2019}, the following functions $\hat{\psi}_A(s)=1/(1+\epsilon^2 s/2D)$ and $\hat{\psi}_M(s)=1/(1+\epsilon^2 s^\alpha/2D_M)$ should be taken into consideration.
The relation between probability $R$ and the absorption coefficient $\kappa$ defined in the system with continuous variables is $R=\kappa^2\epsilon^2/2$. 

Let us assume that the molecule is in region $A$ initially, such that the initial conditions are $P_{A,0}(m;m_0)=\delta_{m,m_0}$ and $P_{M,0}(m;m_0)=0$. After some calculations we get (details are presented in Appendix I)
\begin{eqnarray}\label{eq10}
\hat{P}_A(x,s;x_0)=\frac{1}{2\sqrt{Ds}}\Bigg[{\rm e}^{-|x-x_0|\sqrt{\frac{s}{D}}}\\
+\frac{\sqrt{\frac{s}{D}}-(1-q_A)\sqrt{\kappa^2+\frac{s^\alpha}{D_M}}}{\sqrt{\frac{s}{D}}+(1-q_A)\sqrt{\kappa^2+\frac{s^\alpha}{D_M}}}\;{\rm e}^{-(2x_1-x-x_0)\sqrt{\frac{s}{D}}}\Bigg],\nonumber
\end{eqnarray}
\begin{eqnarray}\label{eq11}
\hat{P}_M(x,s;x_0)=\frac{(1-q_A)s^{\alpha-1}}{D_M\left(\sqrt{\frac{s}{D}}+(1-q_A)\sqrt{\kappa^2+\frac{s^\alpha}{D_M}}\right)}\\
\times\;{\rm e}^{-(x_1-x_0)\sqrt{\frac{s}{D}}-(x-x_1)\sqrt{\kappa^2+\frac{s^\alpha}{D_M}}}.\nonumber
\end{eqnarray}
The Laplace transforms of diffusive fluxes read 
\begin{equation}\label{eq12}
\hat{J}_A(x,s;x_0)=-D\frac{\partial \hat{P}_A(x,s;x_0)}{\partial x},
\end{equation}
\begin{equation}\label{eq13}
\hat{J}_M(x,s;x_0)=-D_M s^{1-\alpha}\frac{\partial \hat{P}_M(x,s;x_0)}{\partial x}.
\end{equation}
Combining the values of the functions Eqs. (\ref{eq10})--(\ref{eq13}) calculated at $x_1$ we get the boundary conditions in terms of the Laplace transform
\begin{equation}\label{eq14}
(1-q_A)D\hat{P}_A(x_1^-,s;x_0)=D_M s^{1-\alpha}\hat{P}_M(x_1^+,s;x_0),
\end{equation}
\begin{equation}\label{eq15}
\hat{J}_A(x_1^-,s;x_0)=\hat{J}_M(x_1^+,s;x_0).
\end{equation}
Using the formula $\mathcal{L}^{-1}[s^\beta \hat{f}(s)]=\partial^\beta f(t)/\partial t^\beta$, $0<\beta<1$, we obtain the boundary conditions in the time domain
\begin{equation}\label{eq16}
(1-q_A)D P_A(x_1^-,t;x_0)=D_M\frac{\partial^{1-\alpha}P_M(x_1^+,t;x_0)}{\partial t^{1-\alpha}},
\end{equation}
\begin{equation}\label{eq17}
J_A(x_1^-,t;x_0)=J_M(x_1^+,t;x_0).
\end{equation}

Assuming that the molecules diffuse independently of one another and all diffusing particles are initially located in the medium $A$, the concentration of molecules can be calculated by means of the formula
\begin{equation}\label{eq18}
C_{A,M}(x,t)=\int_{-\infty}^{x_1} P_{A,M}(x,t;x_0) C_A(x_0,0)dx_0.
\end{equation}
Due to Eq. (\ref{eq18}) the boundary condition for the function $P$ and concentration $C$ are the same. In a similar way, we can derive the boundary conditions at the point $x_2$.
Then, the boundary conditions at both biofilm boundaries are
\begin{equation}\label{eq19}
(1-q_A)D C_A(x_1^-,t)=D_M\frac{\partial^{1-\alpha} C_M(x_1^+,t)}{\partial t^{1-\alpha}},
\end{equation}
\begin{equation}\label{eq20}
J_A(x_1^-,t)=J_M(x_1^+,t),
\end{equation}
\begin{equation}\label{eq21}
D_M\frac{\partial^{1-\alpha} C_M(x_2^-,t)}{\partial t^{1-\alpha}}=(1-q_B)D C_B(x_2^+,t),
\end{equation}
\begin{equation}\label{eq22}
J_M(x_2^-,t)=J_B(x_2^+,t).
\end{equation}
Thus, the diffusive flux is continuous at the boundaries between the media, and the concentration at the boundary in the diffusive medium depends on the concentration in the biofilm at previous times. Such an ageing behaviour is not surprising in the naturally non--stationary scenario of fractional diffusion, equivalent to a Continuous Time Random Walk with diverging $\left\langle t\right\rangle$ \cite{sbm,mjcb}. However, when normal diffusion occurs in the biofilm, the boundary conditions (\ref{eq19}) and (\ref{eq21}) assume a fixed ratio of concentrations at each biofilm boundary.

\section{Theoretical results\label{secIV}}

In the following, we consider a system in which at the initial moment there is a homogeneous solution of antibiotic in the part $A$, while in the other parts of the system there is no antibiotic.
The boundary conditions (\ref{eq19})--(\ref{eq22}) are used to solve equations (\ref{eq1})--(\ref{eq3}) for the following initial condition 
\begin{equation}\label{eq23}
  \left\{\begin{array}{l}
  C_A(x,0)=C_0\;,\\
	C_M(x,0)=0\;,\\
  C_B(x,0)=0\;. 
  \end{array}\right.
\end{equation}
We are interested in calculating the time evolution of the amount of antibiotic $W_B$ that has diffused through the biofilm to region $B$, 
\begin{equation}\label{eq24}
W_B(t)=\Pi\int_{x_2}^\infty C_B(x,t)dx,
\end{equation}
where $\Pi$ is the area of a biofilm surface. 
The function $W_B$ is the basis for our further consideration. 
Below we present the function (\ref{eq24}) in the long time limit. The form of this function depends on the parameter $\kappa$. Details of the calculations are shown in Appendix II.

\subsection{The case of $\kappa=0$\label{secIVA}}

For $\kappa= 0$ we obtain
\begin{equation}\label{eq25}
W_{0B}(t)=C_0 \Pi\left(a_0\sqrt{t}-b_0t^{1-\alpha}\right),
\end{equation}
where 
\begin{equation}\label{eq26}
a_0=\frac{2(1-q_A)\sqrt{D}}{(2-q_A-q_B)\sqrt{\pi}},
\end{equation}
\begin{equation}\label{eq27}
b_0=a^2_0\frac{\pi d(1-q_B)}{2D_M\Gamma(2-\alpha)},
\end{equation}
$d=x_2-x_1$.

\subsection{The case of $\kappa=const.\neq 0$\label{secIVB}}

Assuming $q_A,q_B\neq 1$, we get for $\kappa\neq 0$ 
\begin{equation}\label{eq28}
W_{\kappa B}(t)=C_0 \Pi\left(a_\kappa -b_\kappa\frac{1}{\sqrt{t}}-c_\kappa\frac{1}{t^\alpha}\right),
\end{equation}
where 
\begin{equation}\label{eq29}
a_\kappa=\frac{1}{(1-q_B)\kappa \;{\rm sinh}(\kappa d)},
\end{equation}
\begin{equation}\label{eq30}
b_\kappa=a_\kappa\frac{\coth(\kappa d)}{\sqrt{\pi D}}\left(\frac{1}{1-q_A}+\frac{1}{1-q_B}\right),
\end{equation}
\begin{equation}\label{eq31}
c_\kappa=a_\kappa\frac{1+\kappa d\coth(\kappa d)}{2\kappa^2 D_M\Gamma(1-\alpha)}.
\end{equation}
The characteristic feature of the function $W_{\kappa B}$ Eq. (\ref{eq28}) is that, unlike the function $W_{0B}$, it reaches a plateau for $t\gg {\rm max}((b_\kappa/a_\kappa)^2, (c_\kappa/a_\kappa)^{1/\alpha})$.

\subsection{Biofilm parameters change over time\label{secIVc}}

The results presented in Secs. \ref{secIVA} and \ref{secIVB} have been obtained assuming that the biofilm parameters are constant. However, when the antibiotic acts on the bacteria a biofilm structure can change and biofilm parameters evolve over time. 
Since in such cases the parameters appearing in the equations and boundary conditions depend on time, the derivation of the function $W_B$ requires additional considerations. However, we postulate the use of a quasistatic approximation. In this approximation, we use functions derived for constant parameters, and then assume that these parameters are certain functions of time. The simplest version of this is the following function defined in the case in which antibiotic absorption occurs and biofilm parameters change over time,
\begin{equation}\label{eq32}
W_{\tilde{\kappa}(t)B}(t)=\rho(t)W_{\kappa B}(t), 
\end{equation}
where $\rho(t)$ is to be determined from experimental data. The parameters $a_\kappa$, $b_\kappa$ and $c_\kappa$ for the function $W_{\tilde{\kappa}(t)B}$ are the same as for $W_{\kappa B}$ Eq. (\ref{eq28}). 

The usefulness of this function is shown in Sec. \ref{secVI}. 
Assuming that $\kappa d\ll 1$, which provides $\sinh(\kappa d)\approx 1/\coth(\kappa d)\approx \kappa d$, the function $W_{\tilde{\kappa}(t)B}(t)$ Eq. (\ref{eq32}) can be obtained from the substitution 
\begin{equation}\label{eq33}
\kappa\rightarrow\frac{\kappa}{\rho(t)},\;1-q_{A,B}\rightarrow(1-q_{A,B})\rho(t),\; D_M\rightarrow D_M \rho^2(t)
\end{equation}
in Eqs. (\ref{eq28})--(\ref{eq31}). The above relations define the temporal evolution of the biofilm parameters if Eq. (\ref{eq32}) holds.

\section{Four--stage model of antibiotic diffusion through a biofilm\label{secV}}

Based on the results presented in Sec. \ref{secIV}, we divide the process of antibiotic diffusion in a biofilm into different stages with respect to the following physical characteristics. First, the process can be with or without absorption. These differences appear to be related to the type of bacterial defence mechanism in the biofilm. Secondly, the process can be `static', without changing any parameters, or `dynamic' when at least one of the biofilm parameters changes over time, what is related to the development of biofilm defence mechanisms. Considering the criteria described above, we propose to distinguish four stages described below in the process of antibiotics diffusion in a biofilm. Moreover, for subdiffusion the process is ageing, i.e. the mean mobility is a decreasing function of time. Moreover, if we start the measurement some time after the antibiotic first enters the biofilm, the measurement depends on the ageing time.

It is important to link the stages with the possible defence mechanisms of bacteria in the biofilm. Although the relation of the defence mechanisms to the stages is not immediately obvious, we give below examples of biophysical interpretations of processes that may occur in each stage. 
We mention here that the absorption is treated as a permanent immobilization or disintegration of a molecule. Formally, this process is equivalent to diffusion with an irreversible reaction. However, if the diffusing antibiotic molecule is immobilized temporarily and may continue to diffuse after some time, we treat this process as diffusion with a reversible reaction. The parameters $\alpha$, $D_M$, $q_A$, $q_B$, and $\kappa$ may change due to changes in the biofilm structure. The stages are defined as follows.\\
  \\
{\bf Stage I.} {\it There is no absorption of the antibiotic in the biofilm and all biofilm parameters do not change over time.}\\
   \\
Examples of processes occurring at this stage are the efflux--pump effect and the diffusion of antibiotic molecules in a biofilm in which rapid bacterial growth has been temporarily inhibited, e.g. by limiting the oxygen or nutrient access to bacteria. In this situation the antibiotic molecules may weakly interact with the bacteria because the antibiotic mainly attacks fast-growing bacteria. The efflux pump causes rapid excretion of antibiotic molecules from bacteria. This process can be treated as a subdiffusion with a `reversible reaction' that is described by the equation
	\begin{eqnarray}\label{eq34}
	\frac{\partial C_M(x,t)}{\partial t}=\tilde{D}_M\frac{\partial^{1-\alpha}}{\partial t^{1-\alpha}}\frac{\partial^{2}C_M(x,t)}{\partial x^{2}},
	\end{eqnarray}
where $1/\tilde{D}_M=p/D_a+(1-p)/D_b$, $D_a$ and $D_b$ are coefficients describing molecule random walk outside and inside the bacteria, respectively, $p$ is the probability that the current location of a molecule is outside the bacteria. The derivation of Eq. (\ref{eq34}) is in Appendix III. We mention here that Eq. (\ref{eq34}) for the normal diffusion case was considered in \cite{crank}.\\
    \\
{\bf Stage II.} {\it There is no absorption of the antibiotic and at least one of the biofilm parameters change over time.}\\
  \\
During the initial period, when the concentration of antibiotic in the biofilm is sub--inhibitory, the defence of bacteria against antibiotics is not strong. Then, the bacteria produce little extracellular polymeric substance (EPS). The concentration of antibiotic in the biofilm increases over time, then the EPS is getting denser, which makes diffusion of antibiotic molecules more difficult. However, the density of EPS does not reach such a high concentration that irreversible retention of the antibiotic molecules is possible.\\
\\
{\bf Stage III.} {\it There is absorption of antibiotics in the biofilm, $\kappa\neq 0$, and biofilm parameters do not change over time.}\\
\\
If absorption of antibiotic molecules appears and the values of the parameters are not changed, it may mean that the absorption is carried out by certain `absorption centers' which have appeared as a defensive effect of the bacteria. It is also possible that the density of EPS has reached a constant, high value and the retention of antibiotic molecules occurs with a constant probability.\\
\\
{\bf Stage IV.} {\it There is absorption of antibiotics in the biofilm and at least one of the biofilm parameters changes over time.}\\
\\
Examples of processes occurring at this stage are:
	(a) The diffusion parameters and the absorption parameter change over time. This effect may be due to the increasing high EPS production by bacteria. The density of mucus is so great that it causes immobilization of antibiotic molecules with increasing probability as well as slowing down diffusion.
	(b) Only the absorption parameter changes, the subdiffusion parameters remain constant. Some `absorbing centres' in bacteria are activated that immobilize or destroy antibiotic molecules. The intensity of this process increases over time as the antibiotic concentration increases. During this time, the production of EPS by the bacteria is not so large and changes in subdiffusion parameters are negligibly small.

The division into stages is determined by various forms of the function $W_B$ which can be measured experimentally. 
Based on the empirical results discussed in \cite{tk} and in Sec. \ref{secVI}, the division of the process into stages is supplemented with the following remark: {\it The order of steps depends on both specific antibiotic and biofilm, moreover some stages may not be observed at all.}
While a form of the function in stages I and III is given by Eqs. (\ref{eq25}) and (\ref{eq28}), respectively, the determination of the function for variable parameters, stages II and IV, requires additional considerations. We have not considered the function $W_B$ for stage II since in the examples considered in the next section, this stage is not observed.

\section{Diffusion of ciprofloxacin and gentamicin through {\it Pseudomonas aeruginosa} and {\it Proteus mirabilis} biofilms\label{secVI}}

\begin{figure}[htb]
\centerline{%
\includegraphics[scale=0.35]{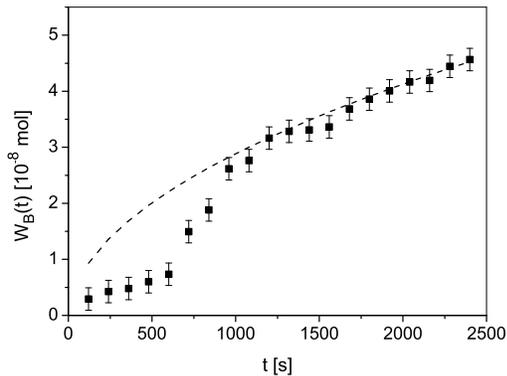}}
\caption{Experimental results (squares) and theoretical function $W_{B0}$ Eq. (\ref{eq25}) (dashed line) for diffusion of ciprofloxacin through {\it P. mirabilis} O18 biofilm, fitting parameters are $a_{0}=0.90\times 10^{-5} \;{\rm m/\sqrt{s}}$ and $b_{0}=0.95\times 10^{-6}\; {\rm m/s^{0.05}}$, and $\alpha=0.95$; here $C_0=1.5\;{\rm mol/m^3}$ and $\Pi=7.0\times 10^{-5}\;{\rm m^2}$.}
\label{fig3}
\end{figure}

\begin{figure}[htb]
\centerline{%
\includegraphics[scale=0.35]{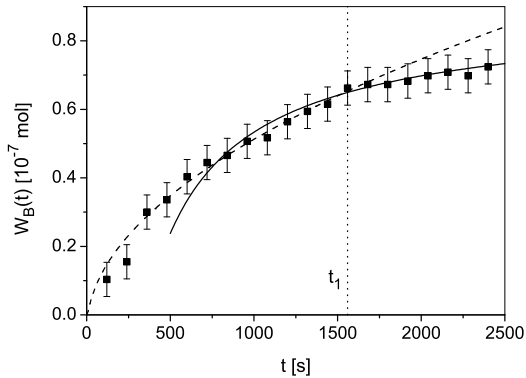}}
\caption{Experimental results (squares) and theoretical functions $W_{0B}$ Eq. (\ref{eq25}) (dashed line) and $W_{\kappa B}$ Eq. (\ref{eq28}) (solid line) for diffusion of ciprofloxacin through {\it Psudomonas aeruginosa} biofilm, the parameters are $a_{0}=0.86\times 10^{-5}\; {\rm m/\sqrt{s}}$, $b_{0}=1.90\times 10^{-5}\; {\rm m/s^{0.05}}$, $a_\kappa=0.44\times 10^{-3}\;{\rm m}$, $b_\kappa=2.10\times 10^{-3}\; {\rm m/\sqrt{s}}$, $c_\kappa=8.57\times 10^{-2}\;{\rm mol/s^{0.95}}$, and $\alpha=0.95$; here $t_1=1560\;{\rm s}$, $C_0=3.0\;{\rm mol/m^3}$, and $\Pi=7.0\times 10^{-5}\;{\rm m^2}$.}
\label{fig4}
\end{figure}

\begin{figure}[htb]
\centerline{%
\includegraphics[scale=0.35]{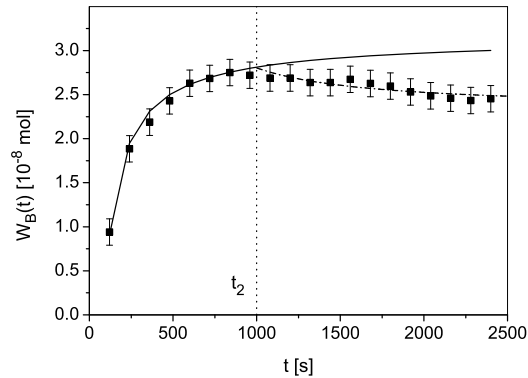}}
\caption{Experimental results (squares) and plots of the functions $W_{\kappa B}$ Eq. (\ref{eq28}) (solid line) and $W_{\tilde{\kappa}(t)B}$ Eq. (\ref{eq35}) (dotted--dashed line) for diffusion of gentamicin in the system with {\it P. mirabilis} O18 biofilm, the parameters are $a_\kappa=0.30\; {\rm 10^{-3} m}$, $b_\kappa=0.43\; {\rm 10^{-3} m/\sqrt{s}}$, $c_\kappa=17.1\;{\rm 10^{-3} m/s^{0.95}}$, $\alpha=0.95$, and $a=1.35$, $b=350 \;{\rm 1/s}$, $t_2=1000\;{\rm s}$; the experiment was performed for $C_0=1.5\; {\rm mol/m^3}$ and $\Pi=7.0\times 10^{-5}\;{\rm m^2}$.}
\label{fig5}
\end{figure}

Diffusion of the antibiotics ciprofloxacin and gentamicin through {\it Pseudomonas aeruginosa} and {\it Proteus mirabilis} biofilms was studied experimentally \cite{ar,ar1}. The experimental setup described in these papers corresponds to the system presented in Fig. \ref{fig1}. At the initial moment, a homogeneous aqueous antibiotic solution (medium $A$) was separated by a biofilm layer (medium $M$) from pure water (medium $B$). For technical reasons, the observation of concentration profiles was possible only in region $B$. Measurements were made in the time interval $\left\langle 100\; {\rm s},2400\; {\rm s}\right\rangle$. Concentration profiles of diffusing substances were measured by means of laser interferometry. Absorption of antibiotic can occur in the biofilm only. Biofilms were cultured on a nucleopore membrane. Since such a membrane is well permeable to antibiotic molecules, we assume that this membrane did not significantly affect the biofilm diffusion properties. The thickness of {\it P. mirabilis} biofilm is $d=5.7\times 10^{-5}\;{\rm m}$. In Figs. \ref{fig3}--\ref{fig5} the experimental data (symbols) and theoretical function $W_B$ (lines) are presented. The experimental data on diffusion of ciprofloxacin through {\it Pseudomonas aeruginosa} PAO1 biofilm were taken from \cite{ar} (presented in Fig. \ref{fig4} in this paper) and the experimental data on diffusion of ciprofloxacin and gentamicin through {\it Proteus mirabilis} O18 biofilm were taken from \cite{ar1} (the data are presented in Figs. \ref{fig3} and \ref{fig5} in this paper).


Analyzing the function $W_B$ obtained experimentally for diffusion of gentamicin through {\it P. mirabilis} O18 biofilm (see Fig. \ref{fig5}), we note that for a long time there persists a stage in which absorption of antibiotic occurs and biofilm parameters change over time. In this case we assume that the function $W_{\tilde{\kappa}(t)B}$ is given by Eq. (\ref{eq32}) with $\rho(t)=1/\left(a-b/t\right)$ for $t>b/a$, where $a$ and $b$ are parameters to be determined. Thus, we get
\begin{equation}\label{eq35}
W_{\tilde{\kappa}(t)B}(t)=\frac{C_0 \Pi}{\left(a-\frac{b}{t}\right)}\left(a_\kappa -b_\kappa\frac{1}{\sqrt{t}}-c_\kappa\frac{1}{t^{\alpha}}\right).
\end{equation}
The parameters $a_\kappa$, $b_\kappa$, and $c_\kappa$ are the same as for the case of $\kappa=const.\neq 0$.

In Figs. \ref{fig3}--\ref{fig5} dashed lines represents the plot of the function $W_{0B}$ Eq. (\ref{eq25}), solid lines represents the plot of $W_{\kappa B}$ Eq. (\ref{eq28}), and dotted--dashed lines are the plots of $W_{\tilde{\kappa}B}$ Eq. (\ref{eq35}). In general, a good agreement between the theoretical functions and the empirical results is observed. In Fig. \ref{fig3} the experimental data on ciprofloxacin diffusion through {\it P. mirabilis} O18 biofilm are well approximated by the function $W_{0B}$ for $t>1000\;{\rm s}$. In Fig. \ref{fig4} the experimental data, presented for the case of ciprofloxacin diffusion through the {\it Pseudomonas aeruginosa} PAO1 biofilm, are well described by $W_{0B}$ for $t<t_1=1560\;{\rm s}$ and by $W_{\kappa B}$ for $t>t_1$. In Fig. \ref{fig5} the functions $W_{\kappa B}$ (for $t<t_2=1000\;{\rm s}$) and $W_{\tilde{\kappa}(t) B}$ (for $t>t_2$) describe the experimental data obtained for gentamicin diffusion through {\it P. mirabilis} O18 biofilm. 

The parameter $\alpha=0.95$ ensures the best fit of theoretical functions Eqs. (\ref{eq25}), (\ref{eq28}), and (\ref{eq35}) to the empirical data. Unfortunately, the empirical data taken from \cite{ar,ar1} do not allow a reliable estimation of the measurement error for this parameter. Because the biofilm constitution is similar to the 1 $\%$ concentration of aqueous agarose solution for which $\alpha=0.95$ \cite{kijhmt}, the assumption that there is subdiffusion in the biofilm seems to be well-justified.

In the time interval $\left\langle 1000\;{\rm s},2400\;{\rm s}\right\rangle$ we observed the stage I only for ciprofloxacin diffusion through {\it P. mirabilis} biofilm (see Fig. \ref{fig3}). For $t<1000\;{\rm s}$, the experimental data are not described by Eq. (\ref{eq25}). 
This is probably due to a finite time needed for antibiotic molecules to pass through the biofilm. In this case we suppose that the bacterial defence mechanisms have not been activated yet. We note that $W_{0B}$ Eqs. (\ref{eq25}) and $W_{\kappa B}$ (\ref{eq28}) have been derived in the limit of long time, so for short time the experimental results may not be described by the functions mentioned above. 
In Fig. \ref{fig3} the function $W_{0B}$ (\ref{eq25}) well describes the experimental data for $t>1000\;{\rm s}$. In the case of gentamicin diffusion through {\it P. aeruginosa} biofilm stages I and III are observed (Fig. \ref{fig4}). The interpretation is that during the initial period $t<1560\;{\rm s}$, when the concentration of the antibiotic in the biofilm is sub--inhibitory, the defence of bacteria against antibiotics is not strong and subdiffusion without absorption with constant biofilm parameters is observed. However, in the next period of time, when the concentration of the antibiotic in the biofilm increases, the antibiotic molecules can be retained or destroyed in the biofilm. Then, bacteria show more active defence against the effects of the antibiotic. Stage III and then stage IV are observed for diffusion of gentamicin through {\it P. mirabilis} biofilm (Fig. \ref{fig5}). In this case, the sub--inhibitory concentration of the antibiotic in the biofilm occurs in a period of time shorter than the time of the first measurement. Activation of the defence mechanisms of bacteria causes that the antibiotic particles are eliminated from the diffusion process initially with a constant probability, and then this probability increases over time, finally reaching a constant value when $t\gg b/a$. According to Eq. (\ref{eq33}), the subdiffusion parameter $D_M$ decreases and the absorption parameter $\kappa$ increases over time. In this stage thickening EPS is probably the dominant bacterial defence mechanism.

The question arises whether subdiffusion or normal diffusion occurs in the biofilm. For the results presented in Figs. \ref{fig3}--\ref{fig5}, the plots of $W_{0B}$ and $W_{\kappa B}$ are best matched with empirical results when $\alpha=0.95$. If the parameter $\alpha$ is less than 1, subdiffusion occurs in the biofilm and the process is described by subdiffusion equation with fractional time derivative. 
\begin{figure}[htb]
\centerline{%
\includegraphics[scale=0.35]{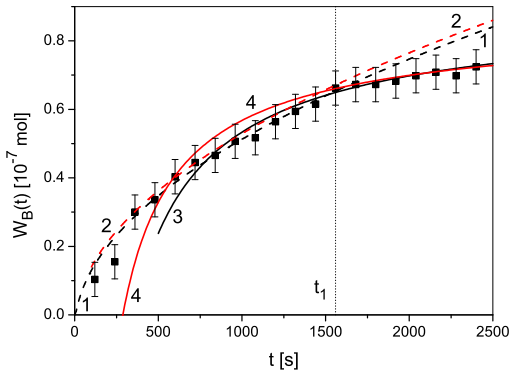}}
\caption{Different stages for the situation of Fig. \ref{fig4}. Dashed lines No. 1 and 2 represent $W_{0B}$ Eq. (25), solid lines No. 3 and 4 represent $W_{\kappa B}$ Eq. (28). Lines No. 1 and 3 are for $\alpha=0.95$, lines No. 2 and 4 are for $\alpha=1.0$. The other parameters are $a_{0}=0.86\times 10^{-5}\; {\rm m/\sqrt{s}}$ and $b_{0}=1.90\times 10^{-5}\; {\rm m/s^{0.05}}$ for the functions No. 1 and 2, $a_\kappa=0.44\times 10^{-3}\;{\rm m}$, $b_\kappa=2.10\times 10^{-3}\; {\rm m/\sqrt{s}}$, and $c_\kappa=8.57\times 10^{-2}\;{\rm mol/s^{0.95}}$ for the function No. 3, and $a_\kappa=0.42\times 10^{-3}\;{\rm m}$ with the same $b_\kappa$ and $c_\kappa$ as in the previous case for the function No. 4.} 
\label{fig6}
\end{figure}
As an example, in Fig. \ref{fig6} we present the plots of theoretical functions obtained for $\alpha=0.95$ and $\alpha=1.0$ for diffusion of ciprofloxacin through {\it Psudomonas aeruginosa} biofilm. We observe a better fit of the theoretical functions to the empirical results for $\alpha=0.95$.

\section{Final remarks\label{secVII}}

We proposed and studied a four-stage model of antibiotic diffusion through a biofilm, along with a possible biological interpretation of the processes occurring in these stages. Subdiffusion of antibiotic molecules may occur in the biofilm, in this case the transport of an antibiotic in a biofilm can be described by the fractional subdiffusion-absorption equation. Physically, this equation describes irretrievable antibiotic molecule immobilisation with power--law sojourn time.
The above conclusions have been obtained by analyzing the temporal evolution of the amount of antibiotic that has diffused through the biofilm $W_B$. Because the function $W_B$ is measurable experimentally, this model gives the opportunity to experimentally check whether absorption occurs in the biofilm and whether the biofilm parameters change over time. 
The course of the process for a particular system depends on the type of antibiotic, its concentration, and the species of biofilm. Not all stages of the process of antibiotic diffusion through the biofilm are always observed. Moreover, in some cases the order of the stages may be different than the one presented in Sec. \ref{secVI} \cite{tk}. 
We emphasize that the experimental measurement is not carried out inside the biofilm, but in the outer region, and is thus non--invasive to the biofilm. The change of biofilm parameters is identified here with the change of parameters of the $W_B$ function. Such physical properties may be useful in deriving new strategies to fight biofilms. We mention that changes in a biofilm structure under the influence of various external factors have been recently intensively studied \cite{dresher,hartmann,diaz,pen}. We believe that knowledge of these facts can be helpful in determining which mechanism of bacterial defence against the effects of an antibiotic dominates the process under consideration. An example of this is diffusion of ciprofloxacin through a {\it Psudomonas aeruginosa} biofilm. This process is presented in Fig. \ref{fig4}, in which ``absorption'' (i.e. the elimination of antibiotic particles from further diffusion in the biofilm) occurs for longer time $t>t_1$. However, it is argued in \cite{swb} that the diffusion barrier should not appear in this case. It can therefore be hypothesized that other biofilm defence mechanisms have been activated that lead to the retention or destroyed of antibiotic molecules. Another hypothesis worth considering is that in this case the diffusion barrier may depend on the concentration of the antibiotic.

If a change in biofilm parameters is observed, it appears likely that the bacteria are actively defending themselves against the effects of the antibiotic. However, if this process is followed by a stage in which the biofilm parameters reach constant values, it probably means that the bacteria do not increase the intensity of their defence despite the fact that the concentration of the antibiotic in the biofilm continues to increase. We therefore hypothesize: {\it If a process with a change in biofilm parameters occurs and a final process is observed in which the parameters are constant when the antibiotic concentration in the biofilm increases, the beginning of the later process is the time at which the bacteria are not able to further enhance an effective defence against the antibiotic using the same defense mechanisms.} A possible biological interpretation is that bacteria were probably killed at that time. For the situation presented in Fig. \ref{fig5}, the final process with constant parameters occurs when the function $W_{\tilde{\kappa}(t)B}$ reaches a plateau.

We suppose that the temporal evolution of antibiotic concentration has the same properties as the function $W_B$. In practice, this means that when calculating antibiotic concentration profiles in a biofilm, one may use the quasistatic approximation in a similar way as it has been done for the $W_B$ function. Considering the diffusion of an antibiotic in a three-dimensional space, the boundary conditions on the biofilm boundary Eqs. (\ref{eq19}) and (\ref{eq20}) can be set in a direction normal to the biofilm surface.

\section*{Appendix I}

The generating functions of Eqs. (\ref{eq5})--(\ref{eq8}) read
\begin{equation}\label{eq36}
S_A(m,z)=\frac{\eta_A^{|m-m_0|}(z)+\Lambda_A(z) \eta_A^{2N-m-m_0}(z)}{\sqrt{1-z^2}},
\end{equation}
\begin{equation}\label{eq37}
S_M(m,z)=\frac{\Lambda_M(z)\eta_M^{m-N-1}(z)\eta_A^{N-m_0}(z)}{\sqrt{(1+Rz)^2-z^2}},
\end{equation}
where $\eta_A(z)=(1-\sqrt{1-z^2})/z$, $\eta_M(z)=(1+Rz-\sqrt{(1+Rz)^2-z^2})/z$, $\Lambda_A(z)=[q-\eta_A(z)+(1-q)\eta_M(z)]/[1/\eta_A(z)-q-(1-q)\eta_M(z)]$, and 
$\Lambda_M(z)=[(1-q)(1-\eta^2_M(z))]/[1/\eta_A(z)-q-(1-q)\eta_M(z)]$.
Moving from discrete to continuous time, we change the variable $z$ to $\hat{\psi}_A(s)$ or $\hat{\psi}_M(s)$ in the generating functions. In \cite{tk2019} there was proved that $\eta_A$ depends on $\hat{\psi}_A(s)$ only, similarly $\eta_M$ depends on the $\hat{\psi}_M$ only. This rule, the equations presented in Sec. \ref{secII} and the approximations $\hat{\psi}_A(s)=1-\epsilon^2 s/2D$, $\hat{\psi}_M(s)=1-\epsilon^2 s^\alpha/2D_M$ provide Eqs. (\ref{eq10}) and (\ref{eq11}) in the limit of small $\epsilon$.

\section*{Appendix II}

The Laplace transforms of solutions to the diffusion equations (\ref{eq1})--(\ref{eq3}) with the boundary conditions (\ref{eq19})--(\ref{eq22}) and the initial condition (\ref{eq23}) are
\begin{eqnarray}\label{eq38a}
\hat{C}_A(x,s)=\frac{C_0}{s}-\frac{C_0 (1-q_A)\beta_M(s)}{s}{\rm e}^{-\beta(s)(x_1-x)}\\
\times\frac{\Xi^+_B(s)+\Xi^-_B(s){\rm e}^{-2\beta_M(s)d}}{\Xi^+_A(s)\Xi^+_B(s)-\Xi^-_A(s)\Xi^-_B(s){\rm e}^{-2\beta_M(s)d}},\nonumber
\end{eqnarray}
\begin{eqnarray}\label{eq38b}
\hat{C}_M(x,s)=\frac{C_0 (1-q_A)D\beta(s)}{s^{2-\alpha}D_M}\\
\times\frac{\Xi^+_B(s){\rm e}^{-\beta_M(s)(x-x_1)}-\Xi^-_B(s){\rm e}^{-\beta_M(s)(2x_2-x_1-x)}}{\Xi^+_A(s)\Xi^+_B(s)-\Xi^-_A(s)\Xi^-_B(s){\rm e}^{-2\beta_M(s)d}},\nonumber
\end{eqnarray}
\begin{eqnarray}\label{eq38}
\hat{C}_B(x,s)=\frac{2C_0 (1-q_A)}{\Xi^+_A(s)\Xi^+_B(s)-\Xi^-_A(s)\Xi^-_B(s){\rm e}^{-2\beta_M(s)d}}\\
\times\frac{\beta(s)\beta_M(s)}{D s} {\rm e}^{-\beta(s)(x-x_2)-\beta_M(s)d},\nonumber
\end{eqnarray}
where $\Xi^\pm_{A,B}(s)=\beta(s)\pm(1-q_{A,B})\beta_M(s)$, $\beta(s)=\sqrt{s/D}$, $\beta_M(s)=\sqrt{\kappa^2+s^\alpha/D_M}$, and $d=x_2-x_1$.
The Laplace transform of time evolution of amount of substance that has diffused through the biofilm is calculated by means of the following formula
\begin{equation}\label{eq39}
\hat{W}_B(s)=\Pi\int_{x_2}^\infty \hat{C}_B(x,s)dx.
\end{equation}
From Eqs. (\ref{eq38}) and (\ref{eq39}) we get
\begin{eqnarray}\label{eq40}
\hat{W}_B(s)=\frac{2(1-q_A)\Pi C_0\beta(s)\beta_M(s)\;{\rm e}^{-\beta_M(s)d}}{(\Xi^+_A(s)\Xi^+_B(s)-\Xi^-_A(s)\Xi^-_B(s){\rm e}^{-2\beta_M(s)d})s\beta(s)}.
\end{eqnarray}
We calculate the inverse Laplace transform in the limit of small $s$, that corresponds to the limit of long time. Keeping the leading terms with respect to $s$ we obtain
\begin{equation}\label{eq41}
  \frac{\hat{W}_B(s)}{\Pi C_0}=\left\{\begin{array}{l}
  \frac{\tilde{a}_0}{s^{3/2}}-\frac{\tilde{b}_0}{s^{2-\alpha}},\;\kappa =0, \\
  \frac{\tilde{a}_\kappa}{s}-\frac{\tilde{b}_\kappa}{\sqrt{s}}-\frac{\tilde{c}_\kappa}{s^{1-\alpha}},\;\kappa\neq 0,
  \end{array}\right.
\end{equation}
where $\tilde{a}_0=(1-q_A)\sqrt{D}/(2-q_A-q_B)$, $\tilde{b}_0=\tilde{a}^2_0 d(1-q_B)/D_M$,
$\tilde{a}_\kappa=1/[(1-q_B)\kappa\sinh(\kappa d)]$, $\tilde{b}_\kappa=\tilde{a}_\kappa\coth(\kappa d)[1/(1-q_A)+1/(1-q_B)]/\sqrt{D}$, $\tilde{c}_\kappa=\tilde{a}_\kappa[1+\kappa d\coth(\kappa d)]/2D_M\kappa^2$.
From Eq. (\ref{eq41}) we get Eqs. (\ref{eq25}) and (\ref{eq28}).

\section*{Appendix III}

In terms of the Laplace transform the general form of the diffusion equation reads \cite{tk2019}
\begin{eqnarray}\label{eq42}
s\hat{P}(x,s;x_0)-P(x,0;x_0)\\
=\frac{\epsilon^2 s\hat{\psi}(s)}{2(1-\hat{\psi}(s))}
\frac{\partial^2\hat{P}(x,s;x_0)}{\partial x^2}.\nonumber
\end{eqnarray}
Let the system consist of two media $a$ and $b$, in which the distributions of waiting time for a next jump are $\hat{\psi}_a(s)=1/(1+\epsilon^2 s^\alpha/2D_a)$ and $\hat{\psi}_b(s)=1/(1+\epsilon^2 s^\alpha/2D_b)$, respectively. The media $a$ and $b$ can be "mixed up" in the system. Currently, the molecule can be in a medium $a$ with probability $p$ or in $b$ with probability $1-p$. The distribution of waiting time for a jump is $\hat{\psi}(s)=p\hat{\psi}_a(s)+(1-p)\hat{\psi}_b(s)$, which for small $s$ reads $\hat{\psi}(s)=1-\epsilon^2(p/2D_a +(1-p)/2D_b) s^\alpha$. Then, from Eq. (\ref{eq42}) we get Eq. (\ref{eq34}).\\

\end{document}